\documentclass[conference]{IEEEtran}
\IEEEoverridecommandlockouts
\usepackage{cite}
\usepackage{amsmath,amssymb,amsfonts}
\usepackage{algorithmic}
\usepackage{graphicx}
\usepackage{textcomp}
\usepackage{xcolor}
\def\BibTeX{{\rm B\kern-.05em{\sc i\kern-.025em b}\kern-.08em
    T\kern-.1667em\lower.7ex\hbox{E}\kern-.125emX}}
\usepackage{tabularx,ragged2e,booktabs,caption}
\usepackage{lipsum}
\ifCLASSINFOpdf
\else
\fi

\begin{document}
%
\title{Multi Time-scale Imputation aided State Estimation in Distribution System}


\author{Shweta~Dahale,~\IEEEmembership{Student Member,~IEEE,}
       Balasubramaniam~Natarajan,~\IEEEmembership{Senior~Member,~IEEE}
\thanks{S. Dahale and B. Natarajan are with Electrical and Computer Engineering, Kansas State University, Manhattan, KS-66506, USA, (e-mail:
sddahale@ksu.edu, bala@ksu.edu)}}


%


\maketitle

\begin{abstract}
With the transition to a smart grid, we are witnessing a significant growth in sensor deployments and smart metering infrastructure in the distribution system. However, information from these sensors and meters are typically unevenly sampled at different time-scales and are incomplete. It is critical to effectively aggregate these information sources for situational awareness. In order to reconcile the heterogeneous multi-scale time-series data, we present a multi-task Gaussian process framework. This framework exploits the spatio-temporal correlation across the time-series data to impute data at any desired time-scale while providing confidence bounds on the imputations. The value of the imputed data for distribution system operation is illustrated via a matrix completion based state estimation strategy. Results on the IEEE 37 bus distribution system reveals the superior performance of the proposed approach relative to linear interpolation approaches.

\end{abstract}


%
\IEEEpeerreviewmaketitle

\section{Introduction}
The goal of state estimation  (SE) is to infer the states of the system based on the measurements collected from it.
Conventionally, SE was limited mostly to the transmission level. The lack of observability at low or medium voltage distribution levels hindered the notion of distribution system state estimation (DSSE). However, with the advent of the smart grid, there has been a plethora of efforts to extract and exploit information at the distribution feeder level. The increase in the deployment of smart meters and sensors have contributed to improved  monitoring of the distribution system. Also, the new generation of PMUs (Phasor Measurement Units) and IEDs (Intelligent Electronic Devices) improve the measurement, protection, and control functions in substations \cite{gomez2011multilevel}. This increase in sensing results in an increase in the information aggregated at centralized distribution management system (DMS). Some researchers envision a hierarchical scheme where data is processed locally \cite{gomez2012state}. With substation level data processing, distributed and localized SE can be implemented. 


\par
The information aggregated from different sensors at the substation level presents some important challenges. Firstly, the data from multiple sources are sampled at different rates and are rarely synchronized. The sources of information available at the distribution substations \cite{gomez2014state} can be broadly classified into two categories: (1) fast rate measurements from RTUs (remote terminal units) sampled at every few seconds and (2) slow rate measurements from smart meters updated every few minutes/hours. Secondly, some of the data may be missing or corrupted due to communication network impairments. The key challenges is to reconcile multiple time series with noisy, heterogeneous, incomplete and unevenly sampled data and estimate the states of the distribution system.
\par
Weighted least squares (WLS) has been the traditional approach for distribution system state estimation (DSSE). In order to guarantee full observability for the WLS-based state estimation, historical data based pseudo-measurements have been used to artificially compensate for insufficient data. These pseudo-measurements do not capture the  real-time spatio-temporal correlations underlying the stochastic environment (e.g., a distribution grid with rooftop PV's).
In \cite{gomez2014state}, the WLS estimation exploits simple linear interpolation and extrapolation technique to reconcile time-series data collected at two different time-scales. However,  this interpolation approach does not exploit the spatio-temporal relationships across the network to impute the time-series data. Secondly, the approach is simplistic and  limited to two time-scale data. Thirdly, the imputed time-series data is still limited only to specific spatial locations in the network where measurements are collected. In practice, there may not be adequate number of measurements to guarantee observability. Hence, the approach in \cite{gomez2014state} is limited in its practical applicability to distribution systems. 
 \par
Recently, the challenge of unobservability in DSSE has been addressed by sparsity-based state estimation approaches. These
approaches exploit the network structure to perform state estimation at low levels of data availability.
Compressive sensing based state estimation technique exploits the sparsity of measurements and states in a linear transformation basis \cite{shafiul}. This technique was further extended for three phase system in \cite{karimi2017compressive}. Matrix completion \cite{donti2019matrix} and tensor completion \cite{madbhavi2020tensor} represent another class of  sparsity-based DSSE techniques that exploit sparsity of raw measurements.  These techniques impute missing elements in the matrix/tensor by obtaining a suitable low rank approximation of the matrix/tensor, respectively. A comprehensive comparison of the performance and complexity associated with these sparsity-based DSSE techniques is presented in \cite{9247106}.
\par
In this paper, we present a new approach for data imputation that leverages the spatio-temporal dependencies in the time series data by using a multi-task Gaussian process (GP) framework. GP's are popular probabilistic models for time-series data built on a non-parametric Bayesian approach \cite{rasmussen2003gaussian}. 
This Bayesian approach not only captures the spatio-temporal correlation across sensor data, it naturally provides uncertainty or confidence estimates associated with the data imputations.
We further leverage these imputed time series data to estimate the voltage states of the distribution system using matrix completion based state estimation technique. 

\subsection{Contributions}
The major contributions of this paper can be summarized as follows:
\begin{enumerate}
    \item A new approach for multi time-scale imputation using multi-task GP framework is proposed. This approach effectively imputes the time-series data in the presence of missing measurements. Error reductions upto 76\% relative to linear interpolation is achieved with 60\% missing measurements. 
    \item The proposed GP-based approach reconciles the time-series by exploiting the spatio-temporal dependencies in the data as opposed to the simple linear interpolation approach in \cite{gomez2014state} that neglects these dependencies.
    \item State estimation using the consistent time-series (resulting from the GP-based approach) is performed using matrix completion under low-observability conditions. The reductions in the error associated with the states is nearly 37\% relative to the estimates relying on linearly interpolated  time-series.   
\end{enumerate}
\par


\section{Sources of Information in smart distribution grid}
\begin{figure}[h!]
\centering
\includegraphics[width = 0.5\textwidth]{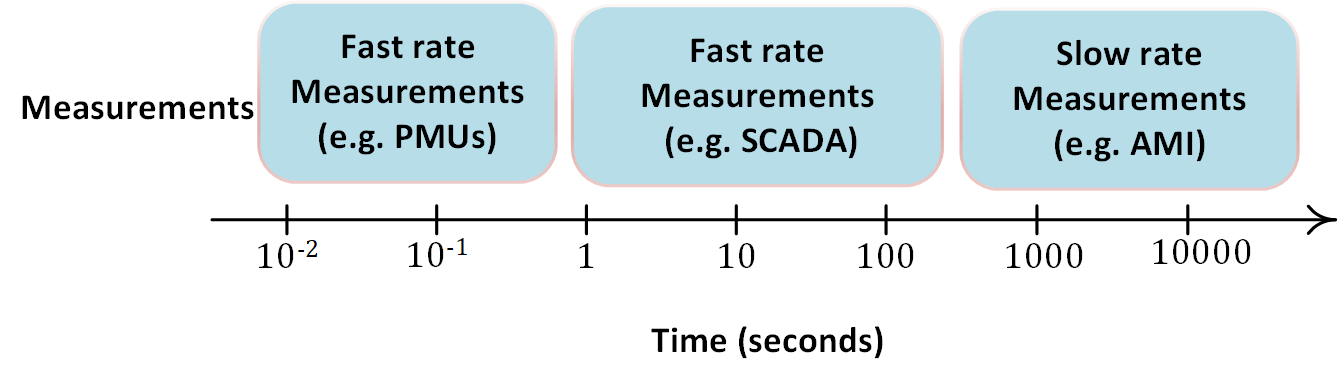}
\caption{Relative frequency of data acquisition from different sources in  distribution systems}
\label{fig:Measurementsources}
\end{figure}

Distribution systems have  poor observability due to insufficient measurements. In order to deal with this challenge, the data collected by multiple sensors at the grid edge could be leveraged. Some of the sources of information available in a smart distribution grid \cite{gomez2014state} are:
\begin{enumerate}
    
    \item Measurements aggregated by the SCADA system that are sampled at rates ranging from a few seconds to about a minute. These measurements typically provide bus voltage and head line currents samples.
    \item MicroPMU data measuring voltage and phase angle at high sampling rates (120 samples/sec for distribution systems).
    \item The power production of  distributed generators needs to be monitored at different rates, ranging from day ahead hourly forecasting to real telemetry periodically transmitted to the DMS.
    \item Advanced Metering Infrastructure (AMI) or smart meter concentrators providing each customers' electricity consumption to the utility at 15 minute or hourly intervals.
    \item Database containing historic load profiles that can be used as pseudo-measurements. While, this data is not as accurate, it improves the observability of the network. 
   
\end{enumerate} 
Fig.\ref{fig:Measurementsources} shows the sources of information available in a smart grid at different timescales \cite{fundamentalnyserda}.
There are two options for real-time DSSE when confronted with multi time-scale measurements \cite{fundamentalnyserda}, namely: (1) perform the DSSE at the resolution of the lowest measurement frequency, or (2) perform the DSSE at a resolution higher than the lowest measurement frequency using a method to reconcile the older and future measurements. For a reliable DSSE, the first option may be unsuitable. In order to pursue the second option, all the information sources should be properly combined as their inclusion may significantly affect the  estimation of system state.


\section{Multi-task GP framework}

\subsection{Background}
A popular approach for multiple timeseries imputation is based upon Gaussian processes. A Gaussian process is a collection of random variables, any finite number of which have a joint Gaussian distribution \cite{rasmussen2003gaussian}. Gaussian process defines a non-parametric prior distribution over functions $$ f(\mathbf{x})  \sim \mathcal{GP}(m(\mathbf{x}), k(\mathbf{x},\mathbf{x}^{'})) $$ 
where, $m(\mathbf{x})$ is the mean function and $k(\mathbf{x},\mathbf{x}^{'})$ is the covariance (kernel) function. The kernel function $k(\cdot) $ dictates the correlation among data points in the modelled function. There are different kernel choices with  one of the most popular being RBF (radial basis function) kernel corresponding to:

\begin{equation}
k(x,x^{'}) = \sigma_s^2 \text{exp}\left(-\dfrac{(x-x^{'})^2}{2l^2} \right)
\end{equation}\\
where hyperparameters $l$ and $\sigma$ are the length-scale and signal variance respectively. These hyperparameters dictate the smoothness of the function. The observed noisy data $y(x)$ has a gaussian distribution with mean $f(x)$ and variance $\sigma^2$, that is $y(x) \sim \mathcal{N}(f(x),\sigma^2)$.
\par
In a GP, the function values  $\mathbf{f_*}$ at test points  $\mathbf{X_*}$ can be computed by conditioning the joint distribution on the observations $\mathbf{X},\mathbf{y}$ \cite{rasmussen2003gaussian}.
Such a  joint distribution of the function $\textbf{f}_{*}$ and observed  outputs $\textbf{y}$ and under the prior is,

\[\begin{pmatrix}
\mathbf{y} \\
\mathbf{f}_{*}
\end{pmatrix}\sim N\left(\begin{pmatrix}
m({\mathbf{X}})\\
m({\mathbf{X}}_*)
\end{pmatrix},\begin{pmatrix}
K(\mathbf{X,X})+\sigma^2 \mathbf{I} & K(\mathbf{X,X_*}) \\
K(\mathbf{X_*,X}) & K(\mathbf{X_*,X_*})
\end{pmatrix}\right).
\]
\\
An extension of GP for multiple signals (tasks) allows us to jointly model and also learn all inter-signal dependencies simultaneously. Please note that we have used the term ``task" to represent ``signal" or a single time-series. A systematic way of multi-task learning is to directly learn a shared covariance
function over tasks.  However, the computational complexity of  computing the inverse of covariance matrix across all data points is $\mathcal{O}(n^3 T^3)$ where $n$ and $T$ are the total number of data points and the total number of parallel time-series, respectively. In order to overcome this computational burden, alternate methods using deep neural networks are formulated in \cite{fortuin2019meta},\cite{fortuin2020gp} and \cite{chung2020deep}. The following section discusses the problem formulation for multi-task GP based data imputation in distribution systems. 

\subsection{Problem Formulation}
Assume we have $P$ sensors at different spatial locations in the distribution grid, each collecting data at a different rate. The training data-set $\mathcal{M}$  consists of $\{\boldsymbol{x_i}, \boldsymbol{y_i}\}_{i=1}^{P}$ where the $i^{th}$ sensor data contains time $\boldsymbol{x_i} \in \mathbb{R}^{n_i \times 1}$ and its corresponding measurement values $\boldsymbol{y_i} \in \mathbb{R}^{n_i \times 1} $. Here, $n_i$ represents the number of observations for the $i^{th}$ sensor data which are different as each sensor collects data at different temporal resolutions.
The training of the GP prior refers to the estimation of the mean and kernel function hyperparameters based on the dataset $\mathcal{M}$. 
\par
Additionally, we have target tasks consisting of time-series data $\mathcal{D}$ = $\{\boldsymbol{\tilde{x}_i}$, $\boldsymbol{\tilde{y}_i}$, $\boldsymbol{\tilde{x}_i}^*,$ $\boldsymbol{\tilde{y}_i}^*\}_{i=1}^{P}$ where 
$\boldsymbol{\tilde{x}_i}$ and $\boldsymbol{\tilde{y}_i}$ are observed data points with their corresponding values respectively. 
The goal of multi-task GP approach is to predict the target task $\{\boldsymbol{\tilde{x}_i}^*, \boldsymbol{\tilde{y}_i}^*\}$  given the observed data and the prior parameters learned from the training dataset.

\subsection{Multi-task learning using Gaussian Process }

As mentioned, multi-task GP enables the imputation of multiple time-series simultaneously  by capturing the spatial and temporal dependencies across them.
The GP prior for multi-task GP is defined as,
\begin{equation}
    f(\cdot) = \mathcal{GP}(m_{\phi}(\cdot), k_{\theta}(\cdot,\cdot)),
    \label{equ2}
\end{equation}
where, $m_{\phi}(\cdot)$ is the mean function and $k_{\theta}(\cdot,\cdot)$ is the co-variance function parameterized by the sets of parameters $\phi$ and $\theta$ respectively.
\\
Function $m_{\phi}(\cdot)$ could be based on a deep neural network (e.g. multilayer perceptron (MLP)) \cite{fortuin2019meta}. In a multi-task paradigm, this function models global trend among the entire set of diverse time-series.
The $\phi$ parameters associated with the mean function are globally shared across all the time-series. $k_{\theta}(\cdot,\cdot)$ is the kernel for the GP prior that captures the temporal correlation within each time-series.
The effective encoding of the prior knowledge about the multiple time-series enables us to use only a few observations/measurements for achieving a high imputation performance.   
\\
The GP prior learnable parameters $\psi$ = \{${{\phi,\theta}}$\}  as defined in (\ref{equ2}) are optimized by maximizing the log marginal likelihood (LML)  given as,
\begin{equation}
\begin{aligned}
\mathbf{{\psi^*}} = & \underset{\mathbf{\psi} }{\text{\enspace argmax}}
& \sum_{i=1}^{P}  \text{log} \enspace  p(\boldsymbol{y_i|x_i},\psi)
\end{aligned}
\label{lml_loss}
\end{equation}
\label{equation5}
where, the log-likelihood is the sum of individual time-series data. An individual log-likelihood can be  computed in closed form as,
$$
\text{log} \enspace  p(\boldsymbol{y_i|x_i},\psi)= \dfrac{-1}{2}\boldsymbol({y_i}-m_{\phi}(\boldsymbol{x_i}))^T (K_{\theta}^{xx} + \sigma^2I)^{-1} $$
$$
(\boldsymbol{y_i}-m_{\phi}({x_i})) - \frac{1}{2} \text{log}|K_{\theta}^{xx} + \sigma^2I|\\
- \frac{n}{2} \text{log}(2\pi) 
$$
This LML is computed by stochastic gradient descent method \cite{fortuin2019meta}.
\\
After the GP prior parameters are optimized and effectively encoded, we predict the target tasks based on observed data. The predictive distribution of $\textbf{\~{y}}_i^*$\ is,

\begin{equation*}
p(\mathbf{\tilde{y}}_i^*|{\tilde{\mathbf{x}}}_i^*,{\tilde{\mathbf{y}}_i},{\tilde{\mathbf{x}}_i)} = \mathcal{N}(\mathbf{m^*, K^*})
\end{equation*}
with
\begin{equation}
\mathbf{m^*} = m_\phi(\tilde{\mathbf{x}}_i^*) + K_{\theta}^{*x} (K_{\theta}^{xx} + \sigma^2I)^{-1} ({\tilde{\mathbf{y}}_i}-m_{\phi}(\mathbf{\tilde{x}}_i))
\label{predmean}
\end{equation}
\begin{equation}
\mathbf{K^*} = K_{\theta}^{**} - K_{\theta}^{*x} (K_{\theta}^{xx} + \sigma^2I)^{-1} (K_{\theta}^{x*})
\label{predvar}
\end{equation}
 The consistent time-series consists of the posterior mean and co-variance as defined in (\ref{predmean}), (\ref{predvar}) computed at those time instances where measurements are unobserved.
 A confidence for credible interval can be computed based on the predicted mean, variance and the desired confidence level.

\begin{figure}[h!]
\centering
\includegraphics[width=0.5\textwidth]{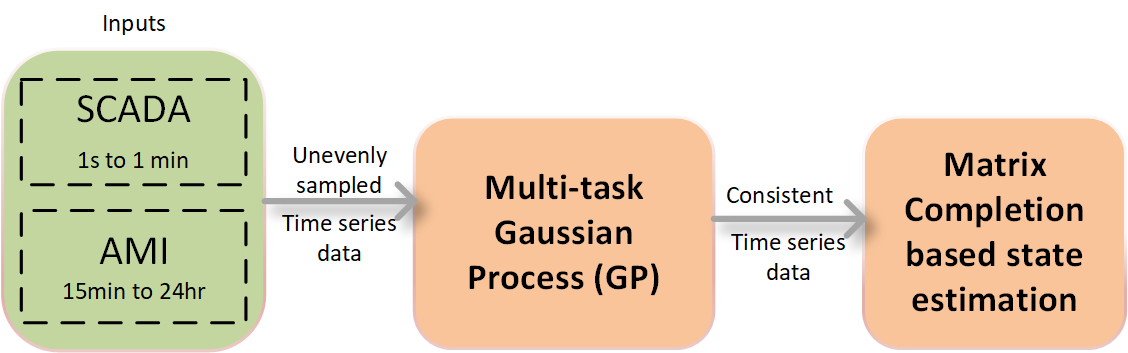}
\caption{Framework for multi time-scale aided state estimation}
\label{fig:Power profile}
\end{figure}
\section{Matrix Completion}
The consistent time-series data from multi-task Gaussian process is still limited to specific spatial locations in the network where sensors are installed. The complete knowledge of the system states are obtained using a matrix completion based state estimation using these limited measurements. \par
Matrix completion aims to determine the incomplete elements in the matrix by obtaining a suitable low rank approximation of the matrix.  
 To utilize the  matrix completion based approach for state estimation, a structured matrix \textbf{X} whose rows correspond to measurement locations and columns correspond to measurement types (e.g. power or voltage) is constructed for a given time. 
  The matrix \textbf{X} is created such that each row represents a bus and each column represents a measurement associated with the bus \cite{9247106}. Therefore, for every bus b $\in$ $B$, the corresponding row in the matrix \textbf{X} $\in$ $\mathbb{R}^{n_1 \times n_2}$
contains:
\begin{equation}
       [\Re(v_b),\Im(v_b ),|v_b |,\Re(s_b ),\Im(s_b )  ] 
       \label{matrixeqn}
\end{equation}
where $n_1 = |B|$ and $n_2$ = 5 quantities per row.  Since the distribution grid is unobservable, \textbf{X} will be an incomplete matrix. The smoothness in spatial variation of the physical measurement quantities (voltage, power, etc) translates into low rank property for this matrix \textbf{X}. Specifically, matrix completion determines the unknown elements
in the matrix by minimizing the nuclear norm of a matrix. More information on matrix completion based DSSE can be found in \cite{9247106}.



\begin{table*}[htbp]
\centering
\caption{Root mean square error of imputed time-series data}
\begin{tabular}{|l|l|l|l|l|l|l|l|l|}
\hline
               & \multicolumn{2}{l|}{\begin{tabular}[c]{@{}l@{}}60\% missing \\      measurements\end{tabular}} & \multicolumn{2}{l|}{\begin{tabular}[c]{@{}l@{}}40\% missing \\     measurements\end{tabular}} & \multicolumn{2}{l|}{\begin{tabular}[c]{@{}l@{}}20\% missing \\    measurements\end{tabular}} & \multicolumn{2}{l|}{\begin{tabular}[c]{@{}l@{}}10\% missing \\    measurements\end{tabular}} \\ \hline
Measurements   & \begin{tabular}[c]{@{}l@{}}Linear \\ Interpolation\end{tabular}             & GP               & \begin{tabular}[c]{@{}l@{}}Linear \\ Interpolation\end{tabular}            & GP               & \begin{tabular}[c]{@{}l@{}}Linear\\ Interpolation\end{tabular}            & GP               & \begin{tabular}[c]{@{}l@{}}Linear \\ Interpolation\end{tabular}           & GP              \\ \hline
Active Power   & 13.28\%                                                                      & 2.6\%              & 5.8\%                                                                      & 1.8\%            & 1.82\%                                                                    & 0.99\%           & 1.5\%                                                                     & 0.7\%           \\ \hline
Reactive Power &  16.8\%                                                                     & 4\%            & 5.3\%                                                                      & 1.4\%            & 1.24\%                                                                    & 1.07\%           & 1.3\%                                                                     & 0.8\%           \\ \hline
\end{tabular}
\label{Tabel1}
\end{table*}

\section{Simulation Results and Discussion}
\begin{figure}[h!]
\centering
\includegraphics[height=6cm, width=9.5cm]{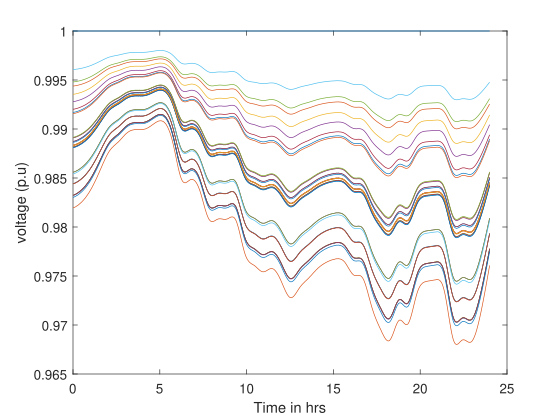}
\caption{Voltage profile}
\label{fig:Voltage profile}
\end{figure}

In this section, the multi-task GP based approach is evaluated on IEEE 37 bus test systems. The loads connected in this system are assumed to comprise of residential homes. The  24-hr active and reactive power consumption profile at load buses for all the phases are synthetically generated. Fig.\ref{fig:Power profile} shows the active power profile of phase A. Reactive power profile is obtained by assuming a power factor of 0.87. A load flow is run in order to generate voltage profiles over the 24-hr period as shown in Fig.\ref{fig:Voltage profile}. As can be seen, the voltage magnitude drops when power consumption increases. This sets the ground truth for imputation.\\
 Snapshots of the AMI measurements (active and reactive power injection) are obtained by sampling the 24-h curves at 15 min interval. SCADA measurements comprising of the voltage magnitudes are sampled at 1 min interval. 
  We split the SCADA and AMI dataset into training and testing data. In order to model the GP prior, a mean function consists of a feed-forward neural network with two hidden layers and ReLu activation functions while the kernel function comprises of RBF kernel. The  hyper-parameters associated with the GP prior are optimized with respect to the log marginal likelihood as given in (\ref{lml_loss}).  Training was performed for 100 epochs using stochastic gradient descent (Adam optimizer).

\begin{figure}[h!]
\centering
\includegraphics[height=6cm, width=10cm]{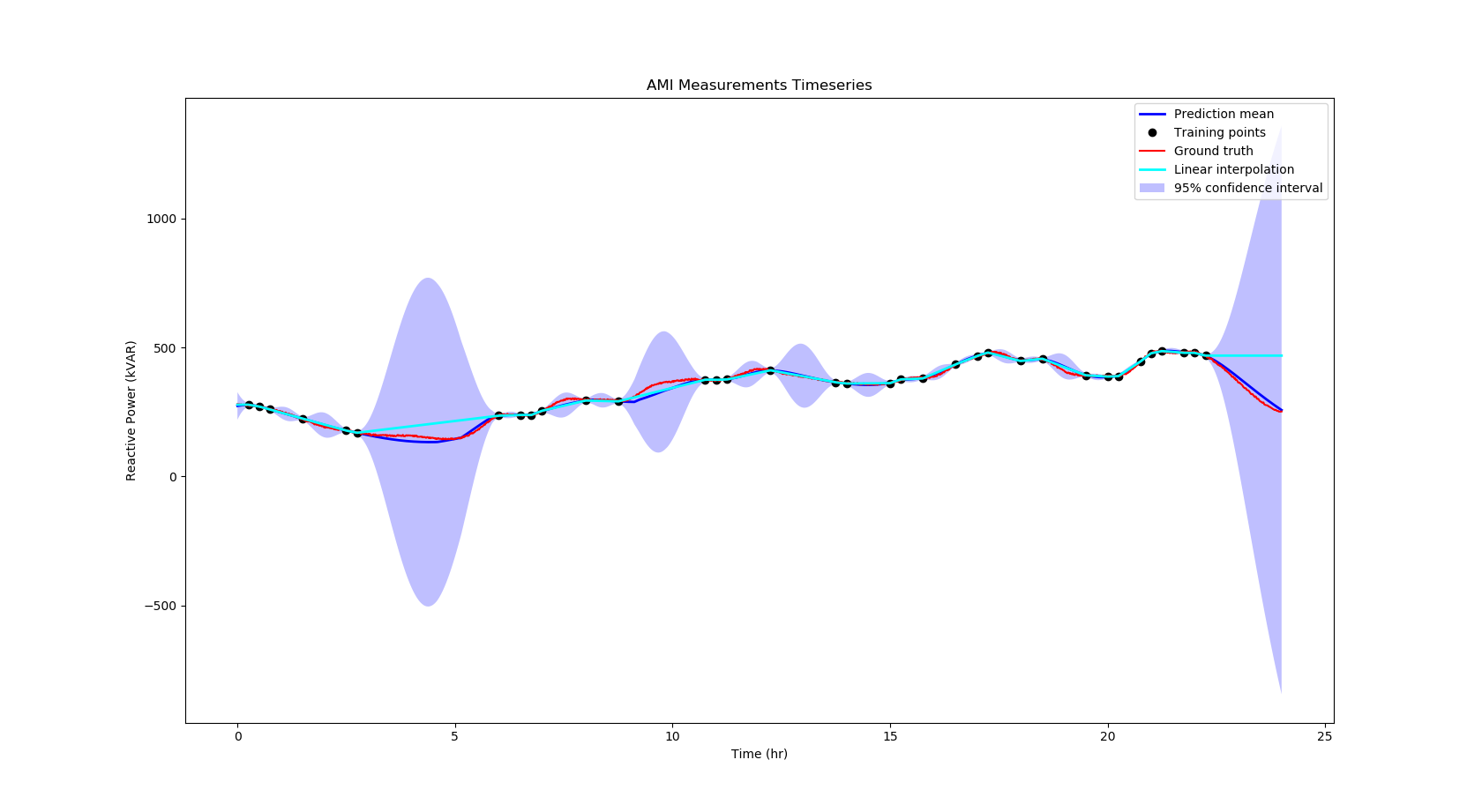}
\caption{Data imputation of AMI timeseries}
\label{fig:AMI data imputation - Task 1}
\end{figure}

\begin{figure}[h!]
\centering
\includegraphics[height=6cm, width=10cm]{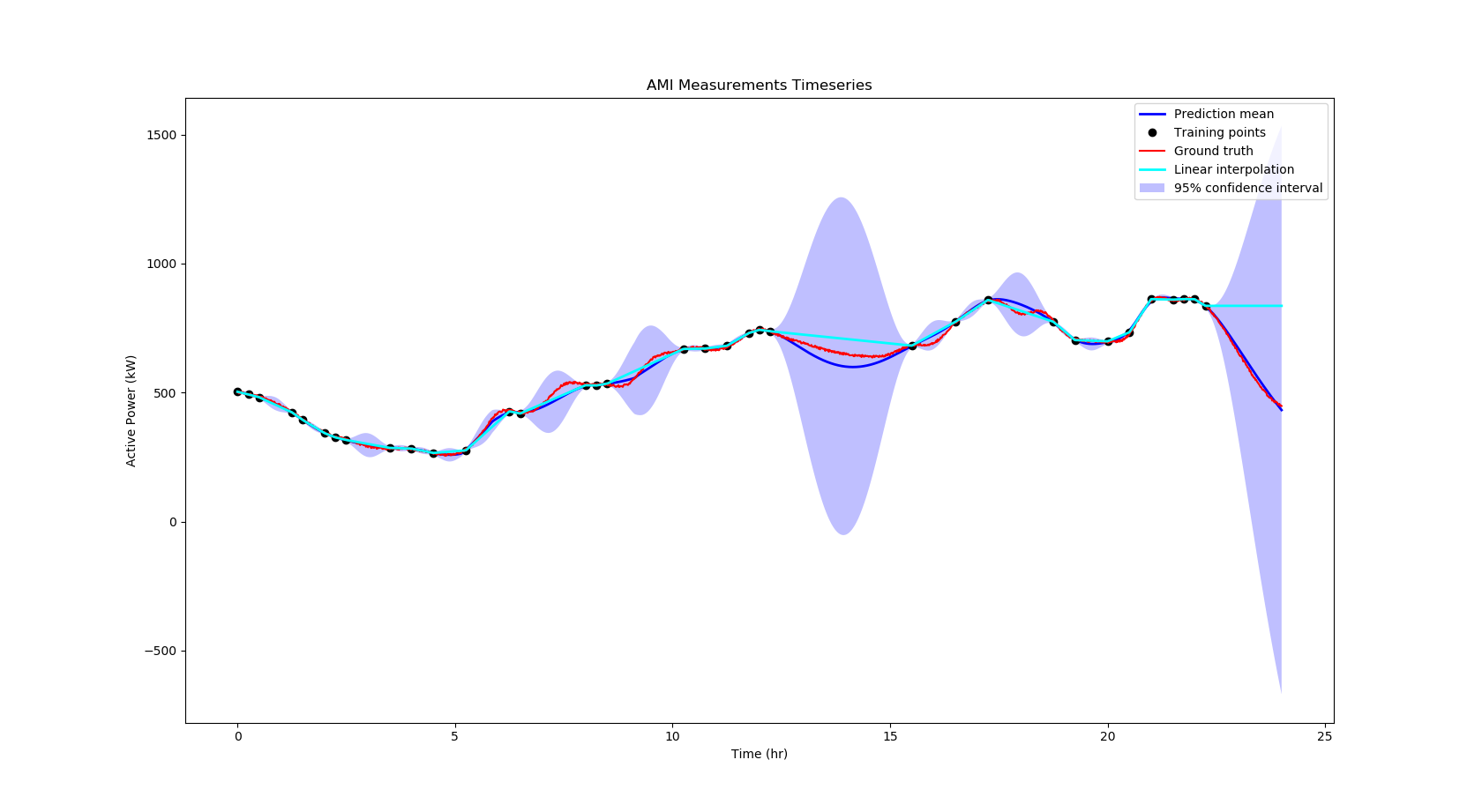}
\caption{Data imputation of AMI timeseries}
\label{fig:AMI data imputation - Task 3}
\end{figure}

\begin{table*}[]
\caption{Absolute errors and relative error reductions (\%) compared to the actual measurements}
\centering
\begin{tabular}{|l|l|l|l|l|l|l|l|l|l|l|l|l|l|l|l|}
\hline
Scenario                & \multicolumn{3}{l|}{FAD = 50\%}           & \multicolumn{3}{l|}{FAD = 60\%}            & \multicolumn{3}{l|}{FAD = 70\%}            & \multicolumn{3}{l|}{FAD = 80\%}            & \multicolumn{3}{l|}{FAD = 90\%}            \\ \hline
Estimated measurements    & \multicolumn{1}{r|}{Linear} & GP   & \%   & \multicolumn{1}{r|}{Linear} & GP    & \%   & \multicolumn{1}{r|}{Linear} & GP    & \%   & \multicolumn{1}{r|}{Linear} & GP    & \%   & \multicolumn{1}{r|}{Linear} & GP     & \%  \\ \hline
Active Power (kW)       & 5.5                         & 5.05 & 8.29 & 4.72                        & 4.34  & 7.8  & 4.01                        & 3.386 & 15.7 & 3.15                        & 2.598 & 17.5 & 2.382                       & 1.61   & 32  \\ \hline
Reactive power (kVAR)   & 2.505                       & 2.5  & 0.2  & 2.097                       & 1.965 & 6.29 & 1.728                       & 1.590 & 7.96 & 1.293                       & 922.9 & 28.6 & 876.3                       & 545.97 & 37  \\ \hline
Voltage magnitude (p.u) & 0.611                       & 0.6  & 1.8  & 0.537                       & 0.501 & 6.7  & 0.437                       & 0.429 & 1.83 & 0.329                       & 0.302 & 8.2  & 0.203                       & 0.19   & 6.4 \\ \hline
\end{tabular}
\end{table*}
   After the GP prior parameters are trained, we perform the imputation for the test data. In this case, we randomly introduce missing measurements over the 24-hr period. Such a  scenario exists in a practical network as AMI and sensors in the network may not transmit data to the utility at all times. Furthermore, the measurements may be lost due to communication network impairments. The unobserved measurements at any instant are imputed at lowest time resolution (here, 1 min interval). Therefore, the fast and slow measurements are reconciled to the same time frame. It is important to note that the GP based imputation can be accomplished in any time-scale. In order to verify the efficacy of the GP-based imputation technique, we compare its performance with the linear interpolation technique presented in \cite{gomez2014state}.
\begin{figure}[h!]
\centering
\includegraphics[height=6cm, width=9.5cm]{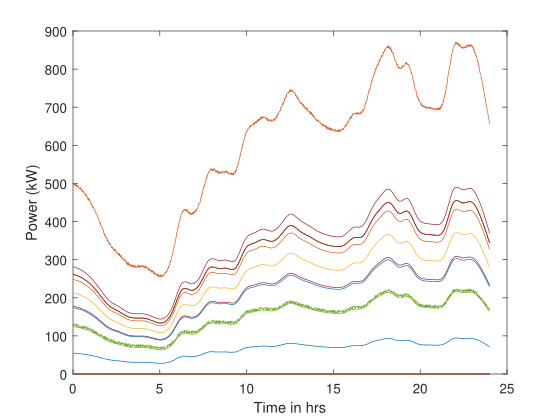}
\caption{Power profile}
\label{fig:Power profile}
\end{figure}
Fig. \ref{fig:AMI data imputation - Task 1} and Fig. \ref{fig:AMI data imputation - Task 3} shows the results  obtained from imputation performed for different AMI profiles  with 60\%  missing measurements. That is, instead of observing power data at regular 15 min interval (96 time instants), we only observe 35 time instants that distributed randomly throughout the 24-hr duration. The uncertainty associated with the missing measurement is indicated by the 95\% confidence interval. As seen from the figures, our approach provides accurate and smoother imputation than linear interpolation technique even with few available measurements. Furthermore, the uncertainty bounds associated with the GP-based imputed data provides a measure of confidence in the measurements. The linear interpolation approach does not provide such confidence bounds.  Table \ref{Tabel1} shows a comparative performance between GP based approach and linear interpolation technique for different percentages of missing measurements. Error reductions of about 76\% is gained from GP-based approach with 60\% missing measurements. This is due to the fact that spatial and temporal correlation in the tasks are exploited by designing optimum mean and kernel hyperparameters. 

 
\par
The consistent time-series data is further used to estimate the states of the system using matrix completion based DSSE. The ratio of the available measurements to the total measurements in the network is indicated by the fraction of available data (FAD). For simulation, FAD is varied from 50\% to 90\% and reconstruction error at each FAD is observed. The absolute errors in the estimated states obtained from both GP as well as linear interpolation time-series data is shown in Table II. It can be inferred that  GP based technique significantly reduces error at all FADs. For example, the  error in estimating reactive power using GP based imputed time-series is reduced by 37\% at 90\% FAD as compared to the linearly interpolated time-series.

\par


\section{Conclusion}
This paper presents a multi-task Gaussian process approach that effectively reconciles data from heterogenous sources in a distribution system by exploiting its spatio-temporal properties. Simulations on the IEEE 37 test system show the superior performance of the proposed approach compared to a linear interpolation technique. The gains in performance are specially pronounced in the presence of missing measurements.  
The consistent time-series data is further used to estimate the system states using matrix completion based DSSE approach. It can be inferred that our proposed GP based approach can be a powerful tool to handle and analyze multi time-scale data in distribution systems.

\section{Acknowledgement}
This material is based upon work supported by the Department  of  Energy,  Office  of  Energy  Efficiency  and  Renewable Energy  (EERE),  Solar  Energy  Technologies  Office,  under Award Number DE-EE0008767.
\bibliographystyle{IEEEtran}
\bibliography{ref.bib}

\end{document}